%Paper: hep-ph/9303276
%From: rebhana@email.tuwien.ac.at (Anton REBHAN)
%Date: Thu, 18 Mar 93 11:40:26 MET

% THE FIGURES ARE CONTAINED IN A PICTEX FILE AT THE END
% OF THIS PLAIN-TEX FILE. IT HAS TO BE EXTRACTED AS A
% SEPARATE TEX FILE SO THAT IT MAY BE TEX-ED. FOR THIS,
% THE MACRO PACKAGE "PICTEX.TEX" IS REQUIRED, WHICH MAY
% BE OBTAINED FROM THE BULLETIN BOARDS BY A GET COMMAND.

\magnification\magstephalf
\tolerance 10000

\overfullrule 0 pt
\font\rfont=cmr10 at 10 true pt
\def\ref#1{$^{{\rfont {[#1]}}}$}

  %%Fonts

\font\twelverm=cmr12
\font\twelvebf=cmbx12
\font\twelveit=cmti12

 %% Toni's
\def\0{\over } \def\1{\vec } \def\2{{1\over2}} \def\4{{1\over4}}
\def\5{\bar } %\def\5{\overline }
\def\6{\partial }
\def\7#1{{#1}\llap{/}}
\def\8#1{{\textstyle{#1}}} \def\9#1{{\bf{#1}}}

\let\a=\alpha \let\b=\beta  \let\d=\delta
\let\e=\varepsilon   \let\th=\theta
 \let\k=\kappa \let\l=\lambda \let\m=\mu
\let\n=\nu  \let\p=\pi \let\r=\rho \let\s=\sigma
  %let\c=\chi \let\ps=\psi
   
  \let\P=\Pi 
  \let\D=\Delta 
\def\mn{{\m\n}}

\def\CD{{\cal D}}

\def\({\left(} \def\){\right)} \def\<{\langle } \def\>{\rangle }
\def\[{\left[} \def\]{\right]}  

\let\el=\eqalign \let\en=\eqno

\def\pmbf#1{\setbox0=\hbox{${#1}$}
        \kern-.025em\copy0\kern-\wd0
        \kern.05em\copy0\kern-\wd0
        \kern-.025em\raise.0433em\box0 }

  %%Greek
\def\a {\alpha} \def\b {\beta}  \def \d {\delta}
\def\e{\epsilon}  
\def\s {\sigma}  
  
\def\pd {\partial}
\def\pmb#1{\setbox0=\hbox{#1}% \kern-.025em\copy0\kern-\wd0
 \kern.05em\copy0\kern-\wd0 \kern-.025em\raise.0433em\box0 }

  %%Fractions

 %

 %%FORMATTING

\def \cl {\centerline}
\parskip=6pt
\parindent=0pt
%\hsize=17truecm\hoffset=-5truemm
%\voffset=-1truecm\vsize=26.5truecm
\def\footnoterule{\kern-3pt
\hrule width 17truecm \kern 2.6pt}

  %%REFERENCES
%     \defref\label{text}
% generates a number, assigns it to \label, generates an entry.
% to call it again: \ref\label
% To list the refs,  \listrefs
% (Extracted and adapted from harvmac.tex by P Ginsparg)

\catcode`\@=11 % This allows us to modify PLAIN macros.

\def\nolabels{\def\wrlabeL##1{}\def\eqlabeL##1{}\def\reflabeL##1{}}
\def\writelabels{\def\wrlabeL##1{\leavevmode\vadjust{\rlap{\smash%
{\line{{\escapechar=` \hfill\rlap{\sevenrm\hskip.03in\string##1}}}}}}}%
\def\eqlabeL##1{{\escapechar-1\rlap{\sevenrm\hskip.05in\string##1}}}%
\def\reflabeL##1{\noexpand\llap{\noexpand\sevenrm\string\string\string##1}}}
\nolabels
\global\newcount\refno \global\refno=1
\newwrite\rfile
\def\defref{$^{{\rfont [\the\refno]}}$\nref}
\def\nref#1{\xdef#1{\the\refno}\writedef{#1\leftbracket#1}%
\ifnum\refno=1\immediate\openout\rfile=refs.tmp\fi
\global\advance\refno by1\chardef\wfile=\rfile\immediate
\write\rfile{\noexpand\item{#1\ }\reflabeL{#1\hskip.31in}\pctsign}\findarg}
%       horrible hack to sidestep tex \write limitation
\def\findarg#1#{\begingroup\obeylines\newlinechar=`\^^M\pass@rg}
{\obeylines\gdef\pass@rg#1{\writ@line\relax #1^^M\hbox{}^^M}%
\gdef\writ@line#1^^M{\expandafter\toks0\expandafter{\striprel@x #1}%
\edef\next{\the\toks0}\ifx\next\em@rk\let\next=\endgroup\else\ifx\next\empty%
\else\immediate\write\wfile{\the\toks0}\fi\let\next=\writ@line\fi\next\relax}}
\def\striprel@x#1{} \def\em@rk{\hbox{}}
\def\lref{\begingroup\obeylines\lr@f}
\def\lr@f#1#2{\gdef#1{\defref#1{#2}}\endgroup\unskip}
\newwrite\lfile
{\escapechar-1\xdef\pctsign{\string\%}\xdef\leftbracket{\string\{}
\xdef\rightbracket{\string\}}}

\def\writestop{\def\writestoppt{\immediate\write\lfile{\string\p
ageno%
\the\pageno\string\startrefs\leftbracket\the\refno\rightbracket%
\string\def\string\secsym\leftbracket\secsym\rightbracket%
\string\secno\the\secno\string\meqno\the\meqno}\immediate\closeout\lfile}}
\def\writestoppt{}\def\writedef#1{}
\catcode`\@=12 % at signs are no longer letters
\def\k{{\bf k}}
\def\P{{\pmbf\Pi}}
\def\nix#1{{}} % suppresses printout

{
\nopagenumbers
{\twelverm
\rightline{hep-ph/9303276}
\rightline{DAMTP 93--06}
\rightline{TUW-93-03}
\vskip 5truecm
\centerline{{\twelvebf THERMALISATION OF LONGITUDINAL GLUONS}}
\bigskip
\bigskip
\cl{P V Landshof{}f}
\smallskip
\cl{{\twelveit
DAMTP, University of Cambridge, England}}
\bigskip
\cl{A Rebhan}
\smallskip
\cl{{\twelveit Inst. f. Theoret. Phys., Tech. Univ. Vienna, Austria}}
\vskip 3 truecm
{\twelvebf Abstract}

In the usual real-time finite-temperature gauge theory both the
physical and the unphysical degrees of freedom are thermalised. We discuss
the alternative approach %, in the only the
%bare propagators that are thermalised are those
%for the physical degrees of freedom.
%That is, the bare propagators of the
%two unphysical components of the gauge field, and of the ghost, remain frozen.
where only the physical transverse components of the gauge field
have bare thermal propagators, whereas the unphysical degrees of
freedom are not heated.
We show how pinch singularities are avoided: sometimes this requires
resummation. If only the hard thermal loop is included in the resummation,
the spatially-longitudinal
component of the gauge field,
which contains an extra collective plasmon mode, becomes fully thermalised,
though the Faddeev-Popov ghost and the remaining unphysical
component of the gauge field remain frozen.
\bigskip
\vfil
March 1993
\eject}}

\pageno=1

%openup 7pt

1. {\bf FREEZING THE UNPHYSICAL DEGREES OF FREEDOM}
\medskip

Finite-temperature field theory in the real-time
formulation\defref\rRT{L V Keldysh, Soviet Physics JETP 20 (1965) 1018;
A Niemi and G W Semenoff, Nucl Phys B230 (1984) 181}
requires a doubling of fields into type 1 and type 2.
All propagators are $2\times2$-matrices and
the information on the thermal distribution of particles
resides in the matrices which diagonalise these propagators.
A bosonic propagator has the form
$$\9D=\9M\9{\tilde D}\9M,                       \en(1.1a)$$
with
$$ \9{\tilde D}=\(\matrix{\D_F&0\cr
0&-\D_F^*\cr}\)                                 \en(1.1b)$$
and
$$\9{M}=\sqrt{n(|K_0|)}\(\matrix{
e^{\b|K_0|/2}& e^{(\s-\b/2)K_0} \cr
e^{(\b/2-\s)K_0}& e^{\b|K_0|/2} \cr}\),         \en(1.1c)$$
where $n(x)=1/(e^{\b x}-1)$ is the Bose-Einstein distribution
function with temperature $T=1/\b$. The constant
$\s$, whose value has to be chosen to be between $0$ and $\b$, parametrises
different versions of the real-time formalism.

In gauge theories, the usual approach\defref\rBHK{C W Bernard, Phys
Rev D9 (1974) 3312;
H Hata and T Kugo, Phys Rev
D21 (1980) 3333}
is to treat physical
and unphysical degrees of freedom on the same footing, which
requires that also the Faddeev-Popov ghost fields are thermalised
(i.e.~have a propagator of the form (1.1)).
In a previous work\defref\PVLAR{P V Landshoff and
A Rebhan, Nucl Phys B383 (1992) 607}
we have shown that, in the real-time
formalism, it is possible and even simpler to thermalise
only the physical transverse
components of the gauge field. Thus only these components involve the
matrix {\bf M} of (1.1c), while the unphysical part of
the gauge boson propagator and the Faddeev-Popov ghost
propagator have the form (1.1a) but with
the matrix
$$\9{M}_0=\(\matrix{
1& e^{\s K_0}\th(-K_0) \cr
e^{-\s K_0}\th(K_0)& 1 \cr}\)                   \en(1.1d)$$
replacing $\9M$.
The value of $\s$  defines the choice of
the time path of the real-time formalism and so must be the same in
both ${\bf M}$ and ${\bf M}_0$. One may choose $\s$ to be some
finite fraction of $\b$, but it
appears to be more natural to make the
original choice of Keldysh, $\s=0$. Then ${\bf M}_0$ is the zero-%
temperature limit of $\9M$ and so the propagators of the unphysical
fields are frozen:
$$\9{M}=\sqrt{n(|K_0|)}\(\matrix{
e^{\b|K_0|/2}& e^{-\b K_0/2} \cr
e^{\b K_0/2}& e^{\b|K_0|/2} \cr}\)~~~~~~~~~~~~~~~
\9{M}_0=\(\matrix{
1& \th (-K_0)\cr
\th (K_0) &1 \cr}\).                    \en(1.1e)$$
We shall use this choice for the rest of this paper.

We adopt the metric $(1,-1,-1,-1)$ and write
$K_\m=(K_0,\k)$,  and  $k_\m=(0,\k)$, so that $k.k=-\k^2$.
If we introduce the spatially-transverse projection operator
$$A_\mn=\d_\mn-{k_\m k_\n \0 k.k}               \en(1.2a)$$
with
$$\d_\mn=g_\mn-g_{\m 0}g_{\n 0}         \en(1.2)$$
the gluon propagator in Feynman gauge takes the form
$${\bf D}^\mn=-A^\mn\;{\bf M\tilde DM}
-(g^\mn-A^\mn)\;\9M_0\9{\tilde DM}_0.           \en(1.3a)$$
In a general linear gauge with quadratic gauge breaking term
${1\02\xi}(A^\m f_\m f_\n A^\n )$, where $f_\mu$ is a
4-vector which is either constant or constructed from derivatives
${\pd / \pd x}$,
one has to replace
$$g_\mn\9{\tilde D}\to \(\matrix{G_\mn\D_F &0\cr
0&-G_\mn^*\D_F^*\cr}\)                          \en(1.3b)$$
with
$$G_\mn(k)=g_\mn-{k_\m \tilde f_\n+\tilde f_\m k_\n\0\tilde f.k}
+(\tilde f^2-\xi k^2){k_\m k_\n\0(\tilde f.k)^2}.       \en(1.3c)$$

In [\PVLAR] we have demonstrated that this formalism
simplifies perturbative
calculations in general gauges, by showing that it reproduces
the known results for the hard thermal part of the
one-loop gluon self-energy and
the two-loop interaction pressure and at the same time
verifying explicitly their complete gauge independence.

However, the bare propagator (1.3a) leads to a potential difficulty
at higher loop orders. Eventually thermal and nonthermal propagators
carrying the same loop momentum are multiplied together
by self-energy insertions,
and the mismatch between the matrices $\9M$ and $\9M_0$ might
give rise to pinch singularities\defref\LvW{N P Landsman
and Ch G van Weert, Phys Rep 145 (1987) 141}\defref\rW{H A Weldon,
Phys Rev D45 (1992) 352} which are absent in the conventional
real-time formalism.
In section 2 we shall inspect the form of
these singularities in general; at least in physical quantities
we expect them to cancel out, which we shall verify in the example
of the 3-loop gluon interaction pressure in section~3.

In recent years it has become clear that perturbation theory
of high-temperature QCD needs to be improved in order to
be able to study physics at the energy scale $g^2T$. For
this, all effects appearing at the scale $gT$, which arise
from the leading temperature contributions of one-loop
diagrams (``hard thermal loops'')\defref\rHTL{O K Kalashnikov
and V V Klimov, Sov J Nucl Phys 31 (1980) 699;
H A Weldon, Phys Rev D26 (1982) 1394;
J Frenkel and J C Taylor, Nucl Phys B334 (1990) 199;
E Braaten and R D Pisarski, Nucl Phys B337 (1990) 569; B339 (1990) 310},
need to be resummed\defref\rPis{E Braaten and R D Pisarski,
Phys Rev Lett 64 (1990) 1338; Phys Rev D42 (1990) 2156;
Phys Rev D46 (1992) 1829}.
%R D Pisarski, Nucl Phys A525 (1991) 175c;
%E Braaten, Nucl Phys B (Proc Suppl) 23 (1991) 351}.
In section 4
we shall consider in particular pinch singularities caused by the
insertion of hard thermal loops into bare gluon propagators,
finding that resummation of hard thermal loops remove these
singularities in the full propagator.

In our formalism the bare propagator separates physical from unphysical modes,
thermalising only the former. However,
the contributions from the hard thermal loops give rise to
a modified spectrum of a gluon plasma: the originally massless
transverse gluons acquire (momentum-dependent) thermal masses.
Moreover there is a new physical
collective mode of {\it spatially longitudinal}
gluons, the plasmon\ref{\rHTL}\defref\Pisres{R D Pisarski,
Physica A158 (1989) 146}.
This is possible because of the
existence of a second projection operator which is transverse
with respect to four dimensions, but spatially longitudinal,
$$B_\mn=g_\mn-{K_\m K_\n\0K^2}-A_\mn.                   \en(1.4)$$
This raises the question whether the division between
physical and unphysical modes in the full propagator can be
as simple as in (1.3a). In fact we shall find in section 4 that
resummation changes
the division of the gluon propagator into thermal and nonthermal
pieces. Indeed, the part of the longitudinal gluon propagator
containing (1.4) also acquires a thermal structure, though  the
unphysical polarisations, which are longitudinal also with
respect to 4 dimensions, remain at zero temperature.

In section 5, we summarise our findings and also comment on
recently reported problems\defref\BKS{R Baier, G Kunstatter
and D Schiff, Phys Rev D45 (1992) R4381} with covariant gauges in the
resummation program of Braaten and Pisarski\ref{\rPis}.

\bigskip
\goodbreak
2. {\bf PINCH SINGULARITIES AT TWO-LOOP ORDER}
\medskip

Consider loop corrections to the gauge-boson propagator (1.3).
The bare propagator (1.3$a$) with
(1.1$e$), corresponding to the Keldysh choice
$\s =0$ of the time path, satisfies
$$D^{11}(K)+D^{22}(K)=D^{12}(K)+D^{21}(K),      \en(2.1)$$
and so, from its definition,  does the full propagator $\CD^{ab}$. Consequently
the self-energy correction
$$\Pi^{ab}=(D^{-1})^{ab}-(\CD^{-1})^{ab}        \en(2.2)        $$
obeys
$$\Pi^{11}(K)+\Pi^{12}(K)+\Pi^{21}(K)
+\Pi^{22}(K)\equiv0.                                    \en(2.3)$$
Furthermore we have $D^{11*}=-D^{22}$ and $D^{12}(K)=D^{21}(-K)$,
which also is valid for the components of $\CD$ and $\Pi$.

In this paper, we shall explicitly work in covariant gauges, though %
it is straightforward to extend  to other gauges that preserve
rotational invariance (with the heat bath at rest). In such gauges the tensor %
structure of the self-energy involves
$$C_\mn={1\0|\k|}(g_{\m 0}-{K_\m K^0\0K^2})K_\n+(\m\leftrightarrow\n),
\qquad D_\mn={K_\m K_\n\0K^2},                           \en(2.4)$$
in addition to the tensors $A_{\mu\nu}$ and $B_\mn$ introduced
in (1.2$a$) and (1.4). $A_\mn$ is orthogonal
to the other three tensors, and all of them are mutually orthogonal
under a trace.

At the one-loop level, the full propagator receives the
contribution $\9D\P\9D$.
In a general covariant gauge, the non-thermal part of the
bare propagator (1.3) has the tensor structure
$G^\mn-A^\mn=B^\mn+\xi D^\mn$.
Because this is orthogonal to $A^\mn$, %
in the above %
product the thermal part of $\9D$, which involves $\9M$, decouples
from the non-thermal part involving $\9M_0$. The term involving the
thermal parts of the two $\9D$'s has the form  %
$$\el{
\9D\P\9D=&\(\matrix{1&1\cr 1&1}\)
\(n(-K_0)\Pi^{12}_a
+n(K_0)\Pi^{21}_a\)\D_F\D_F^*\cr
&+\D_F^2(\ldots)+\D_F^{*2}(\ldots),\cr}                 \en(2.5)$$
where the suffix $a$ denotes the coefficient of $A_{\mu\nu}$ in the tensor
decomposition of $\Pi$, and we have made use of the identity (2.3).
The term involving the non-thermal part of $\9D$ has a similar form,
but with the zero-temperature limit $-\theta(\mp K_0)$
of the Bose distributions $n(\pm K_0)$.

The presence of a term proportional to $\D_F\D_F^*$ leads %
to an ill-defined singular expression, commonly called a
pinch singularity \ref\LvW,
when it appears within a two-loop diagram
and is integrated over $K$,
unless its coefficient vanishes at $K^2=0$,
the pole of $\D_F$.
Usually in the real-time formalism there is %
in addition to the trivial identity (2.3) a relation
containing information on the thermal nature of the self-energy,
$$n(-K_0)\Pi^{12}+
n(K_0)\Pi^{21}=0                                \en(2.6)$$
which removes the pinch. But in our formalism some of the internal
propagators in $\Pi$ correspond to temperature $\b ^{-1}$ and others
to temperature 0, so that the identity (2.6) is not valid.
If we define an effective (generally momentum-dependent) %
temperature associated with the self energy by
$$\b_\Pi={1\0K_0}\ln{\Pi^{21}\0\Pi^{12}}  \en(2.7)$$
then $\Pi^{ab}$ may be diagonalised
according to
$$\P=\9M_{\Pi}^{-1}\(\matrix{\Pi&0\cr0&-\Pi^*\cr}\)\9M_{\Pi}^{-1}, \en(2.8)$$
with $\9M_{\Pi}$ as in (1.1c) but with $\b=\b_\Pi$,
and the potential pinch singularity in (2.5) is determined by
$$\9D\P\9D\bigg|_{\D\D^*}=\(\matrix{1&1\cr 1&1}\)
2i\e(K_0)\[n_\Pi(K_0)-n_D(K_0)\]{\rm Im\,}\Pi   \en(2.5')$$
($n_D(K_0)$ being either $n(K_0)$ or $-\th(-K_0)$). %
So it is present unless Im $\Pi$ vanishes at $K^2=0$.
Explicit calculations show that it does not vanish
-- in a high-temperature expansion there is a leading
term proportional to $\th(-K^2)T^2$; the subleading term is
nonvanishing also for $K^2>0$ and even diverges logarithmically
when $K^2\to0$.
In [\rW] this problem has been discussed in the context of
physical, nonthermalised constituents of a plasma
interacting with thermalised ones.\footnote{$^1$}{%Our % we have now sigma=0
We disagree with the analogous
expressions given in [\rW] because there it was
erroneously assumed that the nonthermal matrix $\9M_0$, introduced
in (1.1d), would be just the identity matrix. This does not
however change any of the conclusions of [\rW].}
With our treatment of unphysical degrees of freedom which
keeps them at zero temperature, there is equally a potential
problem with pinch singularities at $\ge2$-loop orders.

Before addressing the problem of how to handle
the ill-defined terms
associated with pinch singularities in general,
we shall verify the expectation
that this problem should disappear for physical quantities
by inspecting the gluon-interaction pressure. There
pinch singularities potentially appear starting at the
3-loop order.

\goodbreak
\bigskip
3. {\bf CANCELLATION OF PINCH SINGULARITIES IN THE 3-LOOP PRESSURE}
\medskip

In the calculation of the pressure
at 3-loop order, the diagrams containing potential
pinch singularities when calculated with the Feynman rules
introduced in section 1 are shown in figure~1.\footnote{$^2$}{Because
of the less singular behaviour of fermionic propagators, more than
two would be required to have a chance of producing pinch singularities.}
We recall that\ref{\LvW} in each diagram at least one vertex should be
of type 1.

The diagrams of figure~1a give rise to
$${\rm tr}\, \Pi^{1a}D^{ab}\Pi^{bc}D^{c1},              \en(3.1)$$
where the trace is over the Lorentz indices and includes integration
over the momentum argument.
The thermal part of the bare gluon propagator is proportional
to the spatially transverse projection operator $A_\mn$.
Because this  is orthogonal to the other transverse projectors $B,C$
and $D$
which complete the tensor basis (see (1.4) and (2.4)),
the thermal propagator times the self-energy
is again proportional to $A_\mn$. This is orthogonal to the
nonthermal tensor structure in (1.3),
hence
there is no contribution when one gluon propagator is thermal and the
other nonthermal.

With both propagators either thermal or nonthermal
we find for the term proportional to $\D_F\D_F^*$ in (3.1)
$${\rm tr}({\rm Re\,}\Pi_{11}-i\,{\rm tanh}(\b_\Pi K_0/2)\,
{\rm Im\,}\Pi_{11})\,P\,
(n_D(-K_0)\Pi_{12}+n_D(K_0)\Pi_{21})\,P\,\D_F\D_F^*,    \en(3.2)$$
where for thermal propagators  $n_D(x)=n(x)$, $P_\mn=A_\mn$,
and for nonthermal ones $n_D(x)=-\th(-x)$, $P_\mn=G_\mn-A_\mn$.
Considering two thermal propagators first, we find that
the projection operators $A_\mn$ make (3.2) proportional
to Im~$A^\mn(K)\Pi_\mn(K)$. This vanishes however at $K^2=0$,
as we show in Appendix A. %
For the case of two nonthermal propagators
we shall consider general covariant gauges, where
$G_\mn(K)=g_\mn-(1-\xi)K_\m K_\n/K^2$ in (1.3).
We shall need the result, also proved in Appendix A, that
$$K^\m\,{\rm Im\,}\Pi_\mn=O(K^2),                               \en(3.3)$$
as well as
$$K^\m K^\n \Pi_\mn^{ab}\equiv0.                        \en(3.4)$$

The simplest case is the term proportional to $(1-\xi)^2$.
This contribution vanishes identically by virtue of (3.4).
For the other terms we introduce the abbreviations
$$\Pi_I(K)={\rm Im\,}\Pi(K),\qquad \Pi'={\rm Re\,}\Pi_{11}
-i\,{\rm tanh}(\b_\Pi K_0/2)\,{\rm Im\,}\Pi_{11}                \en(3.5)$$
(suppressing Lorentz indices).
The term proportional to $(1-\xi)$ has the structure
$$K^\m\, \Pi_{I\mn}(g-A)^{\n\l}\,\Pi'_{\l\r}K^\r/K^2 \equiv
K^\m\, \Pi_{I\m0}\,\Pi'_{0\r}\,K^\r/\k^2,                       \en(3.6)$$
because of (3.4), which then vanishes at $K^2=0$ by (3.3).
Finally, the term corresponding to Feynman-gauge propagators is
proportional to
$${\rm tr}(g-A)\,\Pi_I(g-A)\,\Pi'=
\Pi_{I00}{K^\m K^\n\0\k^2}\Pi'_\mn+O(K^2),              \en(3.7)$$
upon using (3.3), which vanishes by (3.4).

The other diagram, figure~1b, has a potential pinch singularity
proportional to the 1-2-component of the ghost self-energy.
The part of the ghost-self energy containing the nonthermal part
of the propagator is innocuous, because this fits to the
nonthermal ghost propagators. The critical term arises from
having the thermal part of the gauge propagator coupled to
the ghost propagator.
In covariant gauges, the latter has the form
$$\int d^4Q K^\m A_\mn(Q)\,(K+Q)^\n \,\d((K+Q)^2)\,\d(Q^2)\ldots,  \en(3.8)$$
which, however, vanishes algebraically at $K^2=0$.

We have thus shown that the 3-loop contributions to the gluon
interaction pressure are free from pinch singularities.
Because the pressure is a physical quantity, this can be
viewed as a further test of the consistency of the Feynman rules
introduced in [\PVLAR].
\goodbreak
\bigskip
4. {\bf RESUMMATION OF PINCH SINGULARITIES} %HARD THERMAL LOOPS}
\medskip

Pinch singularities make their appearance in
off-shell Green's functions, for example in the 2-loop
self-energy diagram obtained by opening one line of a 1-loop
subdiagram in figure~1a.
In this section we shall first %
restrict our attention
to the leading temperature corrections of self-energy insertions,
$\Pi\propto g^2T^2$, the so-called hard thermal loop. When we
calculate the latter in our formalism, we obtain
exactly the same result as usually, of the form
$$\P^{\rm HTL}_\mn=\9M^{-1}\(
\tilde\P_A A_\mn+\tilde\P_B B_\mn\)\9M^{-1},    \en(4.1)$$
where $\9M$ is the same matrix as in the spatially transverse piece
of the bare propagator (1.3), and $\tilde\P$ is of a form
analogous to (1.1b).
Pinch singularities are caused by
the second term in (4.1), because it does not match the matrix
$\9M_0$ in the nonthermal part of the bare propagator and because
Im~$\Pi_B$ does not vanish at $K^2=0$ if the light cone is approached
from the spacelike side. These pinch singularities become worse
at higher loop orders, i.e.~the more self-energy graphs are
inserted between bare propagators.

We shall now demonstrate that it is only the perturbative loop
expansion which is ill-defined here, and that the problem disappears
upon resummation of hard thermal loops.
Resummation means that we have to invert
$\CD^{-1}=\9D^{-1}-\P$. In our formalism the gluon propagator
in a general covariant gauge reads
$$\9D^{-1}_\mn=-\9M^{-1}A_\mn \9{\tilde D}^{-1}\9M^{-1}
-\9M_0^{-1}\(B_\mn+{1\0\xi}D_\mn\)\9{\tilde D}^{-1}\9M_0^{-1}.  \en(4.2)$$
(Strictly speaking, the tensors $B_\mn$ and $D_\mn$ have to be
given a matrix form analogous to (1.1b), but for ease of presentation
we shall pretend that they are real quantities.)
The full inverse propagator can now be written as
$$\el{\CD^{-1}=&-\9M^{-1}\[
A_\mn(\9{\tilde D}^{-1}+\tilde\P_A)
+B_\mn(\9M\9M_0^{-1}\9{\tilde D}^{-1}\9M_0^{-1}\9M+\tilde\P_B)\]\9M^{-1}\cr
&-{1\0\xi}\9M_0^{-1}\9{\tilde D}^{-1}\9M_0^{-1}.\cr}            \en(4.3)$$

Inverting (4.3) simply means inverting the individual matrices
appearing as coefficients of the tensors $A$, $B$, and $D$,
where the interesting term is the spatially longitudinal one,
proportional to $B_\mn$. In momentum space the latter reads
$$\CD^{-1}_B=\9M^{-1}
\(\matrix{K^2-\Pi_B+i\e & 2i\e e^{-\b K_0}\th(K_0)\cr
2i\e e^{\b K_0}\th(-K_0) & -K^2+\Pi_B^*+i\e\cr}\)\9M^{-1}.      \en(4.4)$$
Inversion of this matrix, which for $K_0>0$ ($<0$) is of
upper (lower) triangular form with infinitesimal off-diagonal
elements, yields
$$\CD_B=\9M\(\matrix{
{1\0K^2-\Pi_B+i\e}
& {2i\e e^{-\b K_0}\th(K_0)\0(K^2-\Pi_B+i\e)(K^2-\Pi_B^*-i\e)}\cr
{2i\e e^{\b K_0}\th(-K_0)\0(K^2-\Pi_B+i\e)(K^2-\Pi_B^*-i\e)}
&-{1\0K^2-\Pi_B^*-i\e}\cr}\)\9M.                        \en(4.5)$$
Evidently, the pinch singularities have disappeared in this
resummed expression. % In the %!
%hard-thermal-loop approximation $\Pi_B$ is real
%for $K^2>0$, and the off-diagonal term
%is a regular expression proportional to $\d(K^2-\Pi_B)$;
%the difference of the corrected propagator and its complex conjugate;
%for $K^2<0$,
%where $\Pi_B$ has an imaginary part, the off-diagonal term can be put
%equal to zero, because then the explicit $i\e$ becomes can be put to zero.

This implies that pinch singularities caused by hard thermal
loop insertions in the bare propagator are just an artefact of
the ordinary loop expansion. In order to check that this is also
the case beyond the hard-thermal-loop level, we shall restrict
ourselves to Landau gauge. In general, the self-energy contains
also nontransverse
terms proportional to $C_\mn$ (the tensor $D_\mn$ is
excluded however by (A.1)) which in gauges
other than Landau gauge spoil the simplicity of the above arguments,
as seen from Appendix B; in Landau gauge such terms are projected out.
The transverse part of the general self-energy is given by
$$\P=\9M_A^{-1}A_\mn\tilde\P_A\9M_A^{-1}+
\9M_B^{-1}B_\mn\tilde\P_B\9M_B^{-1},            \en(4.6)$$
where for general momentum the effective
temperature associated with $M_A$ or $M_B$,
which is defined by (2.7),
may be different from the physical one and from zero,
depending on the gauge (which means that it does not
have a direct physical interpretation). %*
%beyond the level of hard thermal loops
We therefore have
$$\el{\CD^{-1}=&-\9M_A^{-1}\[
A_\mn(\9M_A\9M^{-1}\9{\tilde D}^{-1}\9M^{-1}\9M_A+\tilde\P_A)
\]\9M_A^{-1}\cr
&-\9M_B^{-1}\[
B_\mn(\9M_B\9M_0^{-1}\9{\tilde D}^{-1}\9M_0^{-1}\9M_B+\tilde\P_B)\]
\9M_B^{-1},\cr}                                 \en(4.7)$$
and again we find that the $2\times2$-structure of the propagator
is
$$\CD_{A,B}=
\9M_{A,B}\(\matrix{
{1\0K^2-\Pi_{A,B}+i\e} & O(\e) \cr O(\e)
&-{1\0K^2-\Pi_{A,B}^*-i\e}\cr}\)\9M_{A,B}.              \en(4.8)$$
Similarly, the Faddeev-Popov ghosts have to be resummed;
only at the level of hard thermal loops there is no
thermal contribution to the ghost self-energy
so that ghosts remain completely nonthermal. From (4.8) we again conclude
that while pinch singularities may arise at higher loop orders, they
are not there in the full theory. They are removed by dressing the
propagators.

\bigskip
5. {\bf DISCUSSION}
\medskip

We have found that upon resummation of hard thermal loops %*
the originally nonthermal part of the
gauge propagator containing spatially longitudinal polarisations
acquires a thermal structure involving the usual matrix $\9M$
given in (1.1e).
This part of the propagator, eq. (4.5),
can be regarded to be thermalised at
the physical temperature for those momenta where one can neglect
the off-diagonal terms in the central matrix on the right-hand side
of (4.5). This is obviously the case when $\Pi_B$ has an imaginary
part, because then the explicit $i\e$ in (4.5) can be put to zero.
The hard-thermal-loop part of $\Pi_B$ reads\ref{\rHTL}
$$\Pi_B=-K^2{m_{el.}^2\0\k^2}\(1-{K_0\02|\k|}\ln{K_0+|\k|\0K_0-|\k|}\)
\equiv K^2 \pi_B,					\en(5.1)$$
where $m^2_{el.}=g^2NT^2/3$,
and this has an imaginary part for space-like momenta, $K^2<0$.
For $K^2>0$, new poles arise from $1/(1-\pi_B)$
corresponding to extra collective (plasmon) modes, %
but for these the explicit $i\e$ is superfluous, too, because
in the full propagator these excitations have finite damping at
the order $g^2T$. It is thus essential to let $\e\to0$ before
performing the high-temperature expansion and restricting to
the hard-thermal loop (5.1), which is real for $K^2>0$.
Therefore only at $K^2=0$, where $\Pi_B$ vanishes,
the $i\e$ in (4.5) still seems necessary.
With it, the explicit off-diagonal
term in (4.5) reads
$$-{2\pi i \d(K^2)e^{-\b K_0} \th(K_0) \0 (1-\pi_B)(1-\pi_B^*)}. \en(5.2)$$
However, $\pi_B$ diverges logarithmically at $K^2=0$ so that
this vanishes unless it gets multiplied by some other factor
that is singular at $K^2=0$.
%is equivalent to zero in the distributional sense.
We expect that this will not happen, because the $K^2=0$ singularity
of the resummed propagator is absent from the start in Coulomb gauge,
see (B.7). %; it is a gauge artefact, not playing any role in the
%calculation of physical, gauge independent quantities.
Since we have found that the explicit off-diagonal terms
in (4.5) can be neglected, %
we conclude that the resummation of hard thermal loops induces
a complete
thermalisation of the originally nonthermal spatially-longitudinal modes
of the gluon. This is physically appealing because it is associated
with the appearance of extra collective modes which have no
counterpart in the bare theory.

Moreover, we have found that the
pinch singularities which seem to be caused by hard thermal loop insertions
at higher loop orders
are just a signal that the bare theory no longer is adequate,
as they go away upon resummation. Hence, they
are an artefact of the loop expansion.

In the conventional formalism, the need to resum hard thermal
loops manifests itself by gauge dependences of one-loop results %
for plasma parameters
at the energy scale $g^2T$, signalling incompleteness of the
unimproved perturbation theory, whereas pinch singularities
are automatically avoided. In our formalism the need to
eliminate pinch singularities provides
another indicator that resummation is needed to obtain complete results.

In fact, as concerns pure gauge modes, our formalism appears to %
be more suitable to
deal with the problems recently encountered in treating
covariant gauges at finite temperature.
In the conventional formalism, even after resummation
the pure gauge modes contained in the gauge
boson propagator
give rise to ambiguous terms in such supposedly physical
quantities as damping rates of collective excitations.
For example, the damping rate of fermionic excitations,
extracted from the dressed fermion self-energy diagram,
is found to be gauge dependent within the class of
covariant gauges\ref{\BKS}, unless an explicit
infra-red regularisation is retained to transform the
singularity of the fermion propagator into a simple
pole\defref\BKSC{A Rebhan, Phys Rev D46 (1992) 4779}.
This complication is due to thermalised unphysical massless modes %*
which aggravate the infrared-behaviour of the fermion
self-energy. In our formalism, which does not thermalise pure gauge modes,
the latter remain nonthermal after resummation of the
transverse hard thermal self-energies, so this problem is
completely avoided. The resummed self-energy is still
gauge dependent, however the leading terms of the imaginary
part of order $g^2T$ are
determined by one-loop diagrams where all lines are thermal.
As we have seen, in covariant gauges the gauge parameter dependent
part of the propagator does not thermalise upon resummation
so that the leading contributions to the resummed damping rates are
manifestly gauge independent.

\goodbreak
\bigskip
{\bf APPENDIX A: \hfil\break
TRANSVERSALITY AND TRACELESSNESS OF ${\rm Im\,}\Pi(K^2=0)$}
\medskip

In this Appendix we give some details of the derivation of
the results used in the text that the
imaginary part of the one-loop self-energy
of gluons in general covariant gauge is both transverse and
traceless at $K^2=0$.

The complete self-energy is double transverse,
$$K^\m K^\n \P_\mn\equiv0               \en(A.1)$$
as a consequence of BRS-invariance, but not necessarily
transverse, $K^\m\P_\mn\not\equiv0$.
Because of (A.1), tracelessness of Im~$\Pi$ at $K^2=0$ implies
$$A^\mn{\rm Im\,}\Pi_\mn\Big|_{K^2=0}=0.        \en(A.2)$$
Im~$\Pi^\m_\m=O(K^2)$ is most easily
checked in Coulomb gauge. Potential gauge
dependences appear in the part of the one-loop diagram
containing both a thermal and a nonthermal propagator.
Because of the form (1.3c), these contain a loop-momentum
vector contracting with a bare 3-vertex, and using tree-level
Ward identities one finds
$${\rm Im\,}\Pi^\m_\m(K)=\int d^4Q \d(Q^2)\d((K+Q)^2)
(1-z^2) \ldots, \qquad z\equiv {\k\9q\0|\k||\9q|},      \en(A.3)$$
but $Q^2=(K+Q)^2=K^2=0$ implies $z^2=1$. Thus (A.2) holds
for all linear gauges. %

We now shall show
$$K^\m{\rm Im\,}\Pi_\mn\Big|_{K^2=0}=0,\qquad {\rm Im\,}\Pi_\mn
\propto \Pi^{11}_\mn,                   \en(A.4)$$
in general covariant gauge. %
For the nonthermal part of $\P_\mn$, transversality holds
in general covariant gauges. We thus have to check
those terms that contain at least one projector $A_\mn$.
They have the form
$$\el{K^\m \Pi^{11}_\mn
&=\int d^4Q \d(Q^2)\d((K+Q)^2)
\[g^{\s\r}Q^2-Q^\s Q^\r\]A_\s^\a(k+q)V_{\a\r\n}(K+Q,-Q,-K)\ldots\cr
&=\int d^4Q \d(Q^2)\d((K+Q)^2)Q^\s A_\s^\a(k+q)
\[g_{\a\n}K^2-K_\a K_\n\]\ldots,\cr}            \en(A.5)$$
where terms with two projectors $A_\mn$ as well as terms
proportional to $(1-\xi)$ have cancelled.
At $K^2=0$ only the last term in (A.5) is kept, which however is found
to equal the integrand in (A.3), hence, vanishes. %

\bigskip
{\bf APPENDIX B: \hfil\break
RESUMMATION OF HARD THERMAL LOOPS IN GENERAL GAUGES}
\medskip

In general gauges that only respect rotational invariance
%which is
%left unbroken by the presence of a %
in the rest frame of the heat bath, the full inverse
propagator (4.3) can be written as
$$\el{\CD^{-1}=&-\9M^{-1}\[
A_\mn\(\9{\tilde D}^{-1}+\tilde\P_A\)
+B_\mn\(\9M\9M_0^{-1}\9{\tilde D}^{-1}(1-g^2)\9M_0^{-1}\9M
+\tilde\P_B\)\]\9M^{-1}\cr
&-\9M_0^{-1}\[C_\mn fg \9{\tilde D}^{-1} +
D_\mn f^2\9{\tilde D}^{-1}\]\9M_0^{-1},\cr}             \en(B.1)$$
where $f$ and $g$ parametrise the possible gauge choices, with $f\not=0$.

The matrices $A$, $B$, $C$, and $D$, which were introduced above,
fulfil
$$\el{ %
& A.B=B.A=A.C=C.A=A.D=D.A=0,\qquad A+B+D=\91,\cr
& A.A=A,\qquad B.B=B,\qquad C.C=-(B+D),\qquad D.D=D,\cr
& (B+D).C=B.C+C.B=C.D+D.C=C. \cr}                       \en(B.2)$$
With these relations one can show that the inverse of a
matrix
$$\9X_\mn=\9a A_\mn+\9b B_\mn+\9c C_\mn+\9d D_\mn,      \en(B.3)$$
if it exists and if it has an analogous decomposition
$$\9X_\mn^{-1}=\5\9a A_\mn+\5\9b B_\mn+\5\9c C_\mn+\5\9d D_\mn, \en(B.4)$$
is given by
$$\el{
&\5\9a=\9a^{-1},\qquad \5\9b=(\9b+\9c\9d^{-1}\9c)^{-1},\cr
&\5\9c=-\9d^{-1}\9c\5\9b,\qquad
\5\9d=\9d^{-1}-(\9d^{-1}\9c)^2\5\9b.\cr}                \en(B.5)$$
(The existence of (B.4) imposes the following conditions on the
commutators of the
matrices in (B.3): $\[\9b,\9d^{-1}\9c\]=0$ and $\[\9d,\9c\9d^{-1}\9c\]
=0$. This is trivially fulfilled by (B.1), because there
$\9c\propto\9d$.)

Applying these formulae to (B.1), we find that the full propagator
resumming the hard thermal loop contributions possesses the
same terms proportional to $A_\mn$ and $B_\mn$, whereas additional
parts with the same thermal structure as obtained in (4.5)
arise, which can be combined with the $B_\mn$-piece in one
tensor $$B_\mn-{g\0f}C_\mn-{g^2\0f^2}D_\mn.		\en(B.6)$$
The only part
remaining nonthermal is $-f^{-2}\9M_0D_\mn\9{\tilde D}\9M_0$,
which vanishes in a homogeneous gauge, where one would scale
$f,g\to\infty$.
In strict Coulomb gauge we have $g/f=K_0/|\k|$, and (B.6) becomes
$$-{K^2\0\k^2}g_{\m0}g_{\n0}				\en(B.7).$$
In this gauge there are no
thermalised Faddeev-Popov ghosts already in the conventional formalism.
We thus end up with the same
resummed quantities as usually in Coulomb gauge\ref{\rPis}.

\medskip\immediate\closeout\rfile\writestoppt
\baselineskip=14pt{{\bf References}}\bigskip{\frenchspacing%
\parindent=20pt\escapechar=` \input refs.tmp\bigskip}\nonfrenchspacing
\vskip 1truecm
{\bf Figure caption}

1 Diagrams for the pressure in 3-loop order that potentially give rise to
pinches.
\bye

%%%%%% PICTEX FILE %%%%%%%%
%\hsize=17truecm\hoffset=-5truemm
%\vsize=25.5 truecm\voffset=-5truemm
\nopagenumbers
\input pictex.tex
 \font\lab=cmr10 at 12 truept
\linethickness=1truemm
\setcoordinatesystem units <1truemm, 1truemm> point at 0 0
\beginpicture

 \lab
\put{(a)} at 0 0

\setsolid
\circulararc 180 degrees from 25 10 center at 25 0
\circulararc 360 degrees from 25 10 center at 30 10
\circulararc 360 degrees from 25 -10 center at 30 -10
\circulararc 180 degrees from 35 -10 center at 35 0

\put{+} at 55 0

\circulararc 180 degrees from 75 10 center at 75 0
\circulararc 360 degrees from 75 10 center at 80 10
\circulararc 180 degrees from 85 -10 center at 85 0

\put{+} at 5 -40

\circulararc 180 degrees from 25 -30 center at 25 -40
\circulararc 360 degrees from 25 -50 center at 30 -50
\circulararc 180 degrees from 35 -50 center at 35 -40

\put{+} at 55 -40

\circulararc 180 degrees from 75 -30 center at 75 -40
\circulararc 180 degrees from 85 -50 center at 85 -40

\put{(b)} at 0 -80
\put{+} at 30 -80

\circulararc 180 degrees from 50 -70 center at 55 -70
\circulararc 180 degrees from 60 -90 center at 55 -90

\setdashes
\circulararc 360 degrees from 75 -10 center at 80 -10

\circulararc 360 degrees from 25 -30 center at 30 -30

\circulararc 360 degrees from 25 10 center at 30 10

\circulararc 360 degrees from 75 -30 center at 80 -30
\circulararc 360 degrees from 75 -50 center at 80 -50

\circulararc 180 degrees from 50 -70 center at 50 -80
\circulararc 140 degrees from 59.7 -68.29 center at 55 -70
\circulararc 140 degrees from 50.3 -91.71 center at 55 -90
\circulararc 180 degrees from 60 -90 center at 60 -80

\put{$\bullet$} at 25 10
\put{$\bullet$} at 75 10
\put{$\bullet$} at 75 -30
\put{$\bullet$} at 25 -30
\put{$\bullet$} at 50 -70
\put{1} at 22 13
\put{1} at 72 13
\put{1} at 72 -27
\put{1} at 22 -27
\put{1} at 47 -67
\endpicture
\rightline{Figure 1}
\bye